\documentclass[sigconf]{acmart}
\usepackage{booktabs} 
\usepackage{color}
\usepackage{comment}
\usepackage{multirow}
\usepackage{algorithm}
\usepackage{algorithmicx,algpseudocode}

\copyrightyear{2017}
\acmYear{2017}
\setcopyright{acmcopyright}
\acmConference{SIGIR '17, August 07-11, 2017}{}{Shinjuku, Tokyo, Japan}
\acmPrice{15.00}
\acmDOI{http://dx.doi.org/10.1145/3077136.3080689}
\acmISBN{978-1-4503-5022-8/17/08}

\begin{document}
\title{AutoSVD++: An Efficient Hybrid Collaborative Filtering Model via Contractive Auto-encoders}

\author{Shuai Zhang}
\orcid{1234-5678-9012}
\affiliation{%
  \institution{University of New South Wales}
  \city{Sydney}
  \state{NSW}
  \postcode{2052}
    \country{Australia}
}
\email{shuai.zhang@student.unsw.edu.au}

\author{Lina Yao}
\affiliation{%
  \institution{University of New South Wales}
  \city{Sydney}
  \state{NSW}
  \postcode{2052}
  \country{Australia}
}
\email{lina.yao@unsw.edu.au}

\author{Xiwei Xu}
\affiliation{%
  \institution{Data61, CSIRO}
  \city{Sydney}
  \state{NSW}
  \postcode{2015}
  \country{Australia}
}
\email{Xiwei.Xu@data61.csiro.au}

\begin{abstract}
Collaborative filtering (CF) has been successfully used to provide users with personalized products and services. However, dealing with the increasing sparseness of user-item matrix still remains a challenge. To tackle such issue, hybrid CF such as combining with content based filtering and leveraging side information of users and items has been extensively studied to enhance performance. However, most of these approaches depend on hand-crafted feature engineering, which is usually noise-prone and biased by different feature extraction and selection schemes. In this paper, we propose a new hybrid model by generalizing contractive auto-encoder paradigm into matrix factorization framework with good scalability and computational efficiency, which jointly models content information as representations of effectiveness and compactness, and leverage implicit user feedback to make accurate recommendations. Extensive experiments conducted over three large-scale real datasets indicate the proposed approach outperforms the compared methods for item recommendation.

\end{abstract}

%
%




\maketitle

\section{Introduction}
With the increasing amounts of online information, recommender systems have been playing more indispensable role in helping people overcome information overload, and boosting sales for e-commerce companies. Among different recommendation strategies, Collaborative Filtering (CF) has been extensively studied due to its effectiveness and efficiency in the past decades. CF learns user's preferences from usage patterns such as user-item historical interactions to make recommendations. However, it still has limitation in dealing with sparse user-item matrix. Hence, hybrid methods have been gaining much attention to tackle such problem by combining content-based and CF-based methods~\cite{Ricci:2010:RSH:1941884}.


However, most of these approaches are either relying hand-crafted advanced feature engineering, or unable to capture the non-triviality and non-linearity hidden in interactions between content information and user-item matrix very well. Recent advances in deep learning have demonstrated its state-of-the-art performance in revolutionizing recommender systems ~\cite{karatzoglou2016recsys}, it has demonstrated the capability of learning more complex abstractions as effective and compact representations in the higher layers, and capture the complex relationships within data. Plenty of research works have been explored on introducing deep learning into recommender systems to boost the performance \cite{salakhutdinov2007restricted, sedhain2015autorec,wang2015collaborative,dziugaite2015neural}. For example, Salakhutdinov et al.~\cite{salakhutdinov2007restricted} applies the restricted Boltzmann Machines (RBM) to model dyadic relationships of collaborative filtering models. Li et al.~\cite{li2015deep} designs a model that combines marginalized denoising stacked auto-encoders with probabilistic matrix factorization.

Although these methods integrate both deep learning and CF techniques, most of them do not thoroughly make use of side information (e.g., implicit feedback), which has been proved to be effective in real-world recommender system \cite{hu2008collaborative, Ricci:2010:RSH:1941884}. In this paper, we propose a hybrid CF model to overcome such aforementioned shortcoming, AutoSVD++, based on contractive auto-encoder paradigm in conjunction with SVD++ to enhance recommendation performance. Compared with previous work in this direction, our contributions of this paper are summarized as follows:

\begin{itemize}
\item Our model naturally leverages CF and auto-encoder framework in a tightly coupled manner with high scalability. The proposed efficient AutoSVD++ algorithm can significantly improve the computation efficiency by grouping users that shares the same implicit feedback together;
\item By integrating the Contractive Auto-encoder, our model can catch the non-trivial and non-linear characteristics from item content information, and effectively learn the semantic representations within a low-dimensional embedding space;
\item Our model effectively makes use of implicit feedback to further improve the accuracy. The experiments demonstrate empirically that our model outperforms the  compared methods for item recommendation.
\end{itemize}

\section{Preliminaries}
Before we dive into the details of our models, we firstly discuss the preliminaries as follows.

\subsection{Problem Definition}

Given user $u = [ 1, ..., N ]$ and item $i = [ 1, ..., M ]$, the rating $r_{ui} \subset R \in \mathbb{R}^{N \times M}$ is provided by user $u$ to item $i$ indicating user's preferences on items, where most entries are missing. Let $\hat{{r_u}_i}$ denote the predicted value of $r_{ui}$, the set of known ratings is represented as $K= \{(u,i)| r_{ui}\ is\ known\}$. The goal is to predict the ratings of a set of items the user might give but has not interacted yet.

\subsection{Latent Factor Models}

\subsubsection{Biased SVD}
Biased SVD~\cite{koren2008factorization} is a latent factor model, unlike conventional matrix factorization model, it is improved by introducing user and item bias terms:
\begin{equation}
\hat{{r_u}_i} = \mu  + b_u + b_i + V_i^T U_u
\end{equation}
where $\mu$ is the global average rating, $b_u$ indicates the observed deviations of user $u$, $b_i$ is the bias term for item $i$,  $U_u \in \mathbb{R}^k$ and $V_i \in \mathbb{R}^k$ represent the latent preference of user $u$ and latent property of item $i$ respectively, $k$ is the dimensionality of latent factor.

\subsubsection{SVD++}
SVD++~\cite{koren2008factorization} is a variant of biased SVD. It extends the biased SVD model by incorporating implicit information. Generally, implicit feedback such as browsing activity and purchasing history, can help indicate user's preference, particular when explicit feedback is not available. Prediction is done by the following rule:
\begin{equation}
\hat{{r_u}_i} = \mu  + b_u + b_i + V_i^T (U_u + |N(u)|^{-\frac{1}{2}} \sum_{j\in N(u)}y_j)
\end{equation}
where $y_j \in \mathbb{R}^f$ is the implicit factor vector. The set $N(u)$ contains the items for which $u$ provided implicit feedback, $N(u)$ can be replaced by $R(u)$ which contains all the items rated by user $u$~\cite{Ricci:2010:RSH:1941884}, as implicit feedback is not always available. The essence here is that users implicitly tells their preference by giving ratings, regardless of how they rate items. Incorporating this kind of implicit information has been proved to enhance accuracy~\cite{koren2008factorization}. This model is flexible to be integrated various kinds of available implicit feedback in practice.

\subsection{Contractive Auto-encoders}
Contractive Auto-encoders (CAE)~\cite{rifai2011contractive} is an effective unsupervised learning algorithm for generating useful feature representations. The learned representations from CAE are robust towards small perturbations around the training points. It achieves that by using the Jacobian norm as regularization:
\begin{equation}
\jmath_{cae}(\theta ) = \sum_{x \in D_n}(L(x,g(f(x))) + \lambda \parallel J_f(x)\parallel_{F}^2)
\end{equation}
where $x\in \mathbb{R}^{d_x}$ is the input, $D_n$ is the training set, $L$ is the reconstruction error, the parameters $\theta = \left \{ W,W^{'},b_h,b_y \right \}$,  $g(f(x))$ is the reconstruction of $x$, where:
\begin{equation}
g(f(x)) = s_g(W^{'}s_f(Wx+b_h) +b_y)
\end{equation}
$s_f$ is a nonlinear activation function, $s_g$ is the decoder's activation function, $b_h \in \mathbb{R}^{d_h}$ and $b_y \in \mathbb{R}^{d_x}$ are bias vectors, $W \in \mathbb{R}^{d_h \times d_x}$ and $W^{'} \in \mathbb{R}^{d_h \times d_x}$ are weight matrixes, same as~\cite{rifai2011contractive}, we define $W=W^{'}$. The network can be trained by stochastic gradient descent algorithm.

\section{Proposed Methodology}

In this section, we introduce our proposed two hybrid models, namely AutoSVD and AutoSVD++, respectively.
\subsection{AutoSVD}
Suppose we have a set of items, each item has many properties or side information, the feature vector of which can be very high-dimensional or even redundant. Traditional latent factor model like SVD is hard to extract non-trivial and non-linear feature representations~\cite{wang2015collaborative}. Instead, we propose to utilize CAE to extract compact and effective feature representations:
\begin{equation}
cae(C_i) = s_f(W \cdot C_i+b_h)
\end{equation}
where $C_i \in \mathbb{R}^{d_c}$ represents the original feature vector, $cae(C_i) \in \mathbb{R}^{k}$ denotes the low-dimensional feature representation.
In order to integrate the CAE into our model, the proposed hybrid model is formulated as follows:
\begin{equation}
\hat{{r_u}_i} = \mu  + b_u + b_i + (\beta  \cdot cae(C_i) + \epsilon_i)^T U_u
\end{equation}
Similar to~\cite{wang2011collaborative}, we divide item latent vector $V_i$ into two parts, one is the feature vector  $cae(C_i)$ extracted from item-based content information, the other part $\epsilon_i  \in \mathbb{R}^{k} (i=1...n)$ denotes the latent item-based offset vector. $\beta$ is a hyper-parameter to normalize $cae(C_i)$ . We can also decompose the user latent vector in a similar way. However, in many real-world systems, user's profiles could be incomplete or unavailable due to privacy concern. Therefore, it is more sensible to only include items side information.

\subsection{AutoSVD++}
While the combination of SVD and contractive auto-encoders is capable of interpreting effective and non-linear feature representations, it is still unable to produce satisfying recommendations with sparse user-item matrix. We further propose a hybrid model atop contractive auto-encoders and SVD++ , which takes the implicit feedback into consideration for dealing with sparsity problem. In many practical situations, recommendation systems should be centered on implicit feedback~\cite{hu2008collaborative}. Same as AutoSVD, we decompose the item latent vectors into two parts. AutoSVD++ is formulated as the following equation:
\begin{equation}
\hat{{r_u}_i} = \mu  + b_u + b_i +(\beta \cdot cae(C_i) + \epsilon_i)^T (U_u + |N(u)|^{-\frac{1}{2}} \sum_{j\in N(u)}y_j)
\end{equation}
Figure 1 illustrates the structure of AutoSVD and AutoSVD++.

\begin{figure}
\includegraphics[width=0.4\textwidth]{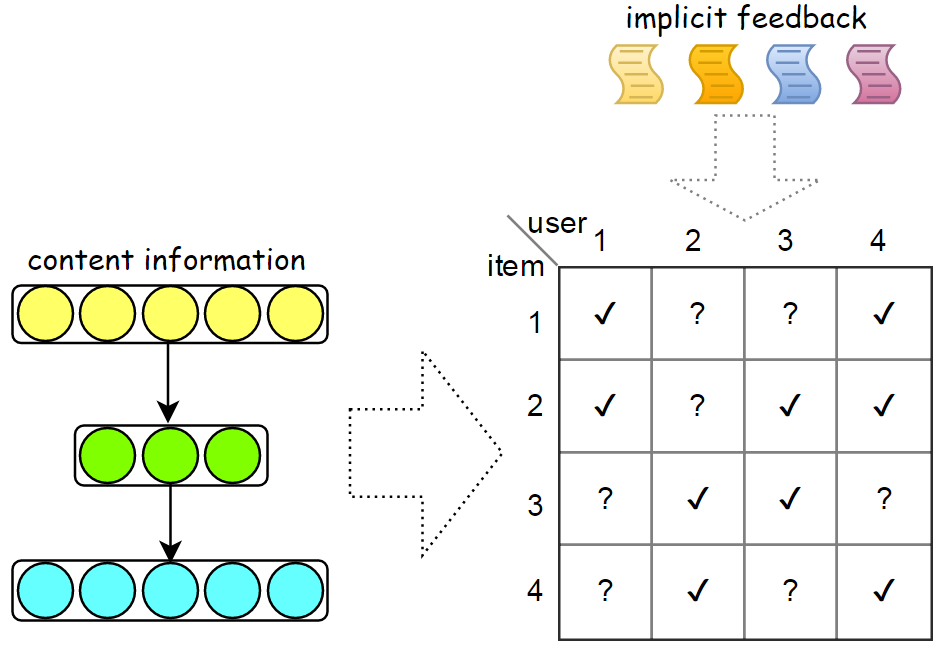}
\caption{Illustration of AutoSVD (remove the implicit feedback) and AutoSVD++.}
\end{figure}

\subsection{Optimization}
We learn the parameters by minimizing the regularized squared error loss on the training set:
\begin{equation}
\underset{b*,\epsilon*,U*,y*}{min}\sum_{\left ( u,i \right ) \in K }(r_{ui} -\hat{r_{ui}})^2 + \lambda \cdot f_{reg}
\end{equation}
where $f_{reg}$ is the regularization terms to prevent overfitting. The $f_{reg}$ for AutoSVD++ is as follows:
\begin{equation}
f_{reg} = b_u^2 +b_i^2 +\left \| \epsilon_i \right \|^2 + \left \| U_u \right \|^2 + \sum_{j\in N(u)}\left \| y_j \right \|^2
\end{equation}
The regularization for AutoSVD is identical to AutoSVD++ with the implicit factor $y_j$ removed.

In this paper, we adopt a sequential optimization approach. We first obtain the high-level feature representations from CAE, and then integrated them into the AutoSVD and AutoSVD++ model. An alternative optimization approach, which optimizes CAE and AutoSVD (AutoSVD++) simultaneously,  could also apply \cite{li2015deep}. However,  the later approach need to recompute all the item content feature vectors when a new item comes, while in the sequential situation, item feature representations only need to be computed once and stored for reuse.

The model parameters are learned by stochastic gradient descent (SGD). First, SGD computes the prediction error:
\begin{equation}
e_{ui} \stackrel{def}{=} r_{ui} -\hat{r_{ui}}
\end{equation}
then modify the parameters by moving in the opposite direction of the gradient. We loop over all known ratings in $K$. Update rules for AutoSVD are as follows:
\begin{align}
 b_u &\leftarrow b_u + \gamma_1  (e_{ui}  - \lambda_1 \cdot  b_u) \label{eq:rel11} \\
 b_i &\leftarrow b_i + \gamma_1  (e_{ui}  - \lambda_1 \cdot  b_i) \label{eq:rel12} \\
 \epsilon_i &\leftarrow \epsilon_i + \gamma_2  (e_{ui}\cdot U_u  - \lambda_2 \cdot  \epsilon_i) \label{eq:rel13} \\
U_u &\leftarrow U_u + \gamma_2  (e_{ui}\cdot(\beta \cdot cae(C_i) + \epsilon_i) - \lambda_2 \cdot U_u) \label{eq:rel14}
\end{align}

Update rules for AutoSVD++ are:
\begin{equation}
\epsilon_i \leftarrow \epsilon_i + \gamma_2  (e_{ui}\cdot(U_u + |N(u)|^{-\frac{1}{2}} \cdot \sum_{j\in N(u)}y_j) - \lambda_2 \cdot  \epsilon_i)
\end{equation}
\begin{equation}
\forall j \in N(u): y_j \leftarrow y_j + \gamma_2 (e_{ui}\cdot |N(u)|^{-\frac{1}{2}}\cdot(\beta \cdot cae(C_i) + \epsilon_i) - \lambda_2  \cdot y_j)
\end{equation}
Where $\gamma_1$ and $\gamma_2$ are the learning rates, $\lambda_1$ and $\lambda_2$ are the regularisation weights. Update rule for $U_u$ of AutoSVD++ is identical to equation $(14)$.

Although AutoSVD++ can easily incorporate implicit information, it's very costly when updating the parameter $y$. To accelerate the training process, similar to  \cite{yang2012local}, we devise an efficient training algorithm, which can significantly decrease the computation time of AutoSVD++ while preserving good performance. The algorithm for AutoSVD++ is shown in Algorithm 1.

\begin{algorithm}
  \caption{Efficient training algorithm for AutoSVD++}\label{AutoSVD++}
  \begin{algorithmic}[1]
    \Procedure{Update Parameters}{}
      \ForAll{user $u$}
        \State $p^{im} \gets |N(u)|^{-\frac{1}{2}} \cdot \sum_{j\in N(u)}y_j$
        \State $p^{old} \gets p^{im}$
        \ForAll{training samples of user $u$}
            \State upadate other parameters
            \State $p^{im} \gets p^{im} + \gamma_2  (e_{ui}\cdot(\beta \cdot  cae(C_i) + \epsilon_i) - \lambda_2 \cdot p^{im})$
        \EndFor
        \ForAll{$i$ in items rated by $u$}
            \State $y_i \gets y_i + |N(u)|^{-\frac{1}{2}} \cdot (p^{im} - p^{old})$
        \EndFor
      \EndFor
    \EndProcedure
  \end{algorithmic}
\end{algorithm}

\section{Experiments}
In this section, extensive experiments are conducted on three real-world datasets to demonstrate the effectiveness of our proposed models.

\subsection{Experimental Setup}
\subsubsection{Dataset Description}
We evaluate the performance of our AutoSVD and AutoSVD++ models on the three public accessible datasets. MovieLens\footnote{https://grouplens.org/datasets/movielens} is a movie rating dataset that has been widely used on evaluating CF algorithms, we use the two stable benchmark datasets, Movielens-100k and Movielens-1M. MovieTweetings\cite{Dooms13crowdrec} is also a new movie rating dataset, however, it is collected from social media, like twitter. It consists of realistic and up-to-date data, and incorporates ratings from twitter users for the most recent and popular movies. Unlike the former two datasets,  the ratings scale of MovieTweetings is 1-10, and it is extremely sparse.
The content information for Movielens-100K consists of genres, years, countries, languages, which are crawled from the IMDB website\footnote{http://www.imdb.com}. For Movielens-1M and Movietweetings, we use genres and years as the content information. The detailed statistics of the three datasets are summarized in Table 1.

\begin{table}
  \caption{Datasets Statistics}
  \label{tab:freq}
  \begin{tabular}{lcccc}
    \toprule
    dataset&\#items&\#users&\#ratings&density(\%)\\
    \midrule
    MovieLens 100k&1682&943&100000&6.30\\
    MovieLens 1M&3706&6040&1000209&4.46\\
    MovieTweetings&27851&48852&603401&0.049\\

  \bottomrule
\end{tabular}
\end{table}

\subsubsection{Evaluation Metrics}
We employ the widely used Root Mean Squared Error (RMSE) as the evaluation metric for measuring the prediction accuracy. It is defined as
\begin{equation}
RMSE = \sqrt{\frac{1}{|T|}\sum_{(u,i) \in T}(\hat{r_{ui}}-r_{ui})^2}
\end{equation}
where $|T|$ is the number of ratings in the testing dataset, $\hat{r_{ui}}$ denotes the predicted ratings for $T$, and $r_{ui}$ is the ground truth.

\subsection{Evaluation Results}
\subsubsection{Overall Comparison}
Except three baseline methods including NMF, PMF and BiasedSVD, four very recent models closely relevant to our work are included in our comparison.

\begin{itemize}
\item \textbf{RBM-CF} \cite{salakhutdinov2007restricted}, RBM-CF is a generative, probabilistic collaborative filtering model based on restricted Boltzmann machines.
\item \textbf{NNMF (3HL)} \cite{dziugaite2015neural}, this model combines a three-layer feed-forward neural
network with the traditional matrix factorization.
\item \textbf{mSDA-CF} \cite{li2015deep} , mSDA-CF is a model that combines PMF with marginalized denoising stacked auto-encoders.
\item \textbf{U-AutoRec} \cite{sedhain2015autorec}, U-AutoRec is novel CF model based on the autoencoder paradigm. Same as~\cite{sedhain2015autorec}, we set the number of hidden units to 500.

\end{itemize}
We use the following hyper-parameter configuration for AutoSVD in this experiment, $\gamma_1 = \gamma_2 = 0.01$, $\lambda_1 = \lambda_2 = 0.1$, $\beta=0.1$ . For AutoSVD++, we set $\gamma_1 = \gamma_2 = 0.007$, $\lambda_1 = 0.005$, $\lambda_2 = 0.015$, and $\beta=0.1$. For all the comprison models, we set the dimension of latent factors $k = 10$ if applicable. We execute each experiment for five times, and take the average RMSE as the result.

According to the evaluation results in Table 2 and Figure 2(a), our proposed model AutoSVD and AutoSVD++ consistently achieve better performance than the baseline and compared recent methods. On the ML-100K dataset, AutoSVD performs slightly better than AutoSVD++, while on the other two datasets, AutoSVD++ outperforms other approaches.

\begin{table}[]
\centering
\caption{Average RMSE for Movielens-100k and Movielens-1M from compared models with different training data percentages.}
\label{my-label}
\begin{tabular}{lcclcl}
\toprule
\multicolumn{1}{c}{Methods} & \multicolumn{2}{c}{ML-100K} & \multicolumn{1}{c}{Methods}              & \multicolumn{2}{c}{ML-1M} \tabularnewline
\multicolumn{1}{c}{}                         & 90\%         & 50\%         & \multicolumn{1}{c}{}                                       & 90\%           &    50\%      \tabularnewline
\midrule
NMF                                          & 0.958        & 0.997        & NMF                                                        & 0.915          &   0.927       \tabularnewline
PMF                                          & 0.952        & 0.977        & PMF                                                        & 0.883          &  0.890        \tabularnewline
NNMF(3HL)                                    & 0.907        & *            & U-AutoRec                                                  & 0.874          &     0.911     \tabularnewline
mSDA-CF                                      & *            & 0.931        & RBM-CF & 0.854         &   0.901
\tabularnewline
Biased SVD                                   & 0.911        & 0.936        & Biased SVD                                                 & 0.876          &    0.889     \tabularnewline
SVD++                                   & 0.913        & 0.938       & SVD++                                                 & 0.855         &  0.884        \\ \hline
AutoSVD                                       & {\bf 0.901 }       & {\bf 0.925 }        & AutoSVD                                                     & 0.864          & 0.877         \tabularnewline
AutoSVD++                                     & 0.904        & 0.926        & AutoSVD++                                                   & {\bf 0.848 }        &  {\bf 0.875  } \\

\bottomrule
\end{tabular}

\end{table}

\subsubsection{Scalability}
Figure 2(b) shows CPU time comparison in log scale.  Compared with traditional SVD++ and Original AutoSVD++, our efficient training algorithm achieves a significant reduction in time complexity. Generally, the optimized AutoSVD++ performs $\bar{R}$ times better than original AutoSVD++, where $\bar{R}$ denotes the average number of items rated by users\cite{yang2012local}. Meanwhile, compared with biased SVD model, the incorporated items $Cae(C_i)$ and offset $\epsilon_i$ does not drag down the training efficiency. This result shows our proposed models are easy to be scaled up over larger datasets without harming the performance and computational cost.

 \begin{figure}[!tb]
\begin{center}
\begin{minipage}[t]{3.8cm}
\includegraphics[width=3.8cm]{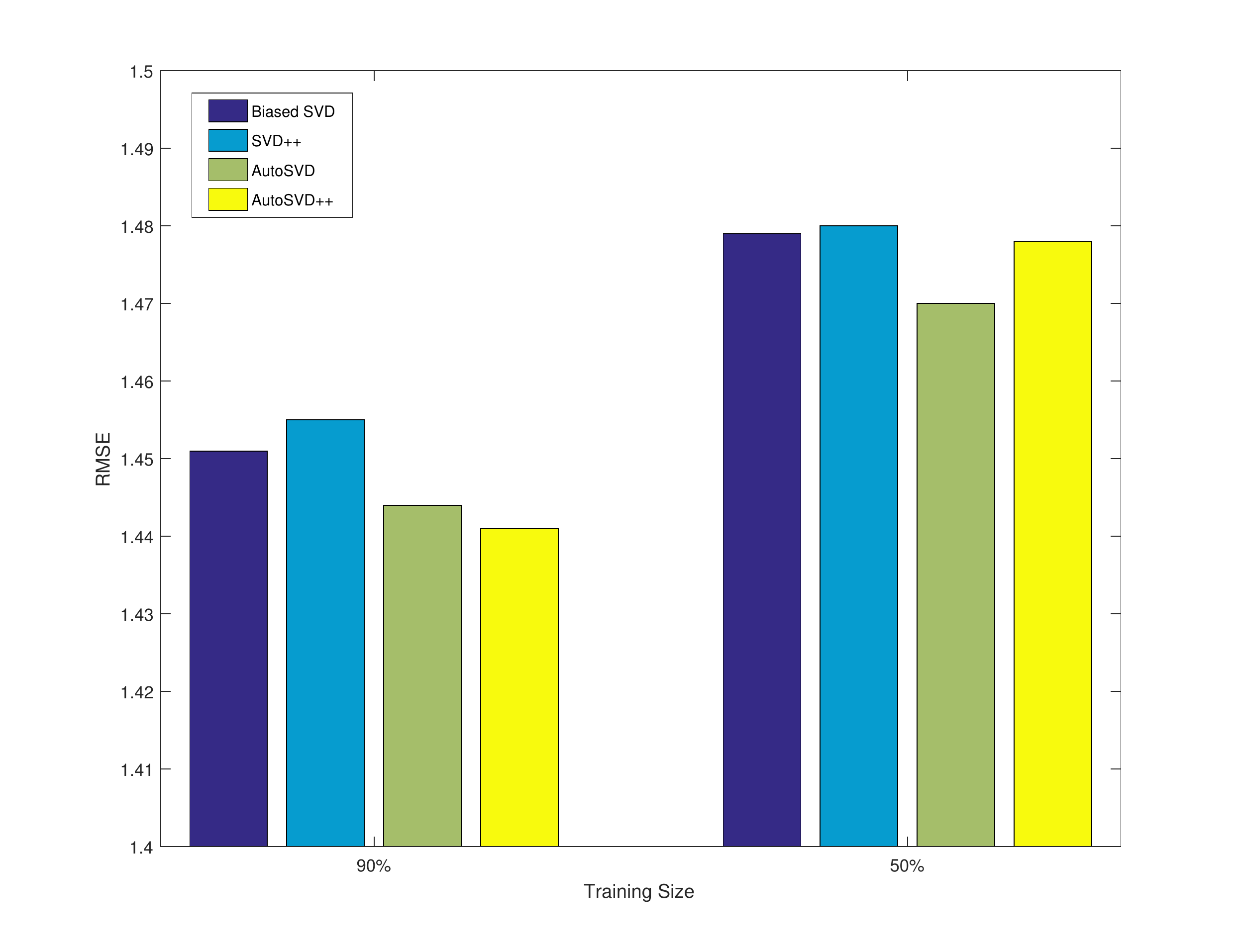}
\centering{(a)}
\end{minipage}
\begin{minipage}[t]{4.6cm}
\includegraphics[width=4.6cm]{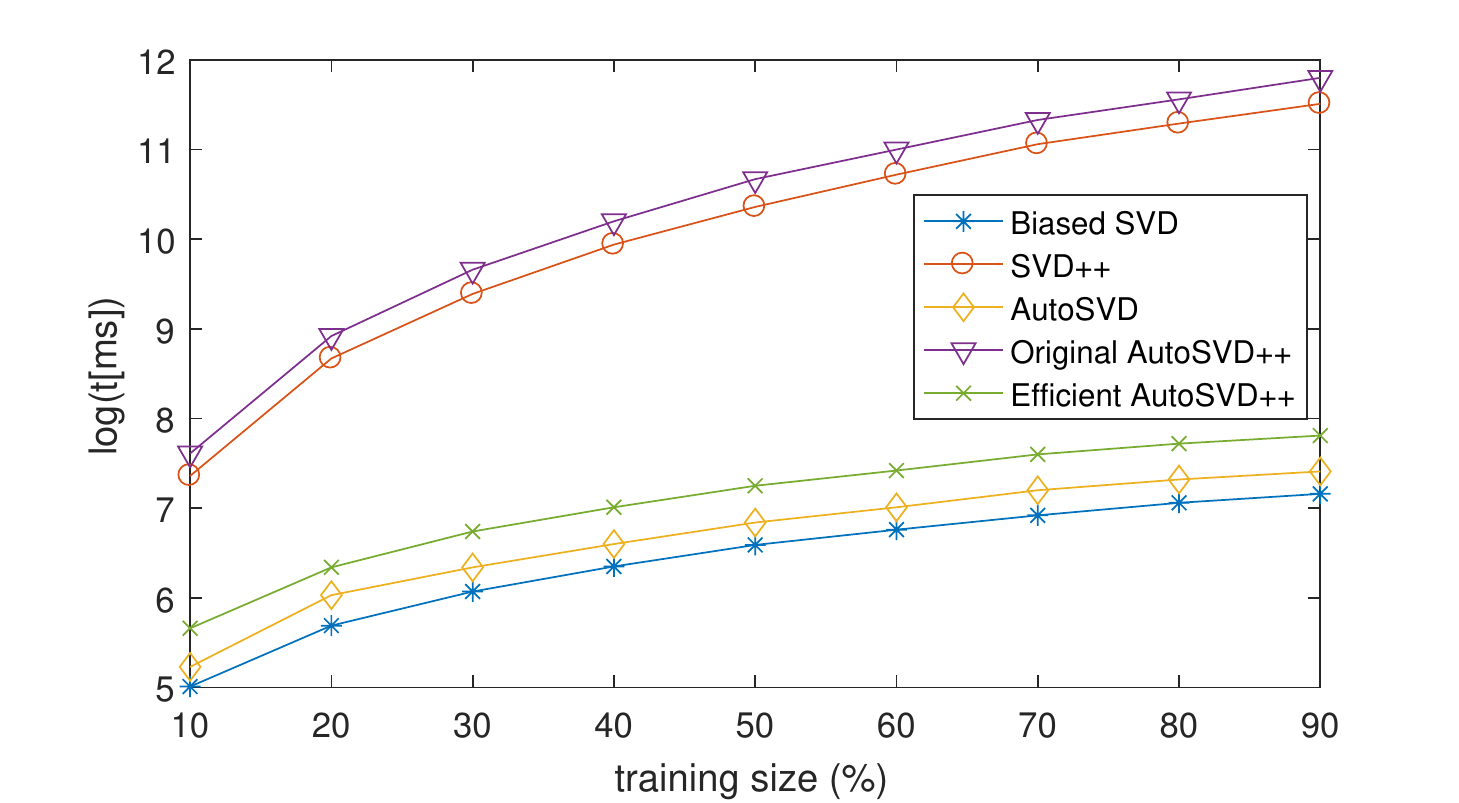}
\centering{(b)}
\end{minipage}
\caption{(a) Average RMSE Comparson on MovieTweetings Dataset (the lower the better). (b) Comparison of training time of one epoch on Movielens-100K (the lower the better).}
\label{fig:observation}
\end{center}
\vspace{-6mm}
\end{figure}

\section{Conclusions And Future Work}
In this paper, we present two efficient hybrid CF models, namely AutoSVD and AutoSVD++. They are able to learn item content representations through CAE, and AutoSVD++ further incorporates the implicit feedback. We devise an efficient algorithm for training AutoSVD++, which significantly speeds up the training process. We conduct a comprehensive set of experiments on three real-world datasets. The results show that our proposed models perform better than the compared recent works.

There are several extensions to our model that we are
currently pursuing .
\begin{itemize}
\item First, we will leverage the abundant item content information such as textual, visual information and obtain richer feature representations through stacked Contractive Auto-encoders;

\item Second, we can further improve the proposed model by incorporating temporal dynamics and social network information.
\end{itemize}

\bibliographystyle{ACM-Reference-Format}
\bibliography{sigproc}

\end{document}